\documentclass[a4paper,12pt]{article}
\usepackage[cp1251]{inputenc}
\usepackage[russian]{babel}
\usepackage{graphics}

\newcommand{\be}{\begin{equation}}
\newcommand{\ee}{\end{equation}}
\begin{document}

{\Large\bf Local dark energy:

HST evidence from the expansion flow

around Cen A/M83 galaxy group}

\vspace{1cm}

{ A.D.~Chernin$^{1,2}$, I.D.~Karachentsev$^3$, D.I.~Makarov$^3$,
O.G. Kashibadze$^3$, P.~Teerikorpi$^2$, M.J.~Valtonen$^{2}$,
V.P.~Dolgachev$^1$, L.M.~Domozhilova$^1$ }

\vspace{1cm} {\it $^1$Sternberg Astronomical Institute, Moscow
University,Moscow, 119899, Russia, e-mail: chernin@sai.msu.ru,

$^2$Tuorla Observatory, Turku University, Piikki\"o, 21 500,
Finland,

$^3$Special Astrophysical Observatory, Nizhnii Arkhys, 369167,
Russia}

\vspace{1cm}

A structure with a massive group in its center and a cool
expansion outflow outside is studied around the Cen A galaxy with
the use of the Hubble Space Telescope observations. It is
demonstrated that the dynamics of the flow is dominated by the
antigravity of the dark energy background. The density of dark
energy in the cell is estimated to be near the global cosmological
density. This agrees with our previous result from the
neighborhood of the Local group. A notion of the ``Hubble cell''
is introduced as a building block of the local structure of the
universe.


\vspace{1cm}

\section{Introduction}

The Hubble flow of expansion is a natural tool to probe and
measure the energy content of the universe. As is well-known, dark
energy first revealed itself in the Hubble magnitude versus
redshift diagram as a weak extra dimming beyond that expected of
the light from  the Type Ia supernovae [1,2] at large redshifts.
The diagram for Ia supernovae showed that, at even larger
redshifts, the effect decreases with its sign changing near $z_V
\simeq   0.7$, so that a weak relative light enhancement beyond
that expected is observed at $z > z_V$. This has reasonably been
taken [1-6] as an indication that the global cosmological
expansion was being decelerated by gravity at times earlier than
$z = z_V$ and accelerated by antigravity at times later than $z =
z_V$.

At the redshift $z_V \simeq 0.75 \pm 0.05$ and at the
corresponding distance $\sim 1000$ Mpc, the antigravity of dark
energy and the gravity of matter (baryons and dark matter) balance
each other for a moment. The balance condition is $\rho_M (z_V) -
2 \rho_V = 0,$ where $\rho_M$ is the matter density, $\rho_V$ is
the dark energy density. It is assumed hereafter that antigravity
is described by the Einstein cosmological constant $\Lambda$, so
that $\rho_V = \Lambda/(8\pi G) > 0$. According to this simple,
straightforward and quite likely interpretation, dark energy is
the energy of the vacuum with the equation of state $p_V = -
\rho_V$ [7]. Here $p_V$ is the dark energy pressure, $G$ is the
gravitational constant and the speed of light $c = 1$. It is also
taken into account that the effective gravitating density of dark
energy (in General Relativity) is $\rho_V + 3 p_V = - 2 \rho_V <
0$.

Since matter density scales with redshift as $(1 + z)^3$ and the
present-day matter density is known, $\rho_M (z=0) \simeq 0.3
\times 10^{-29}$ g cm$^{-3}$, the estimate of the dark energy
density comes from Eq. (1), if the redshift value $z_V$ is found
in observations:
\be \rho_V = \frac{1}{2}\rho_M (z=0) (1 + z_V)^3 \simeq (0.75 \pm
0.05) \times 10^{-29} g/ cm^{3}.\ee \noindent This figure has
later been confirmed by the CMB anisotropy studies with WMAP
observations [8,9].

A rather similar procedure enabled us to detect dark energy on
relatively  small spatial scales and estimate its density from the
distance-redshift diagram for the local expansion flow in our
close galactic neighborhood [10-13]. High accuracy observations of
the local galaxies' flow with the Hubble Space Telescope [14-28]
were used to discover local antigravity and estimate the dark
energy density at a few Mpc from the Milky Way. The local density
of dark energy proved to be near, if not exactly equal to, the
global figure of Eq.1. This result is independent of, compatible
with, and complementary to large-distance observations.

In this paper, we report on the study of the local outflow around
the galaxy group known as the Cen A/M83 complex.  The data on the
complex and its vicinity are used which have come from original
observations with the Hubble Space Telescope [29]. Local dark
energy effects are in the focus of the study. We demonstrate that
the dynamical structure of the flow is significantly affected by
the antigravity of the dark energy background. The local density
of dark energy is estimated to be near the global cosmological
density, or exactly equal to it. In Sec.2, the notion of the
Hubble cell is introduced and theory considerations are outlined
about local flows of expansion dominated by dark energy; in Sec.3,
the basic data on the outflow around the galaxy complex Cen A/M83
are presented and analyzed; in Sec.4 the results are summarized.

\section{Local expansion and dark energy}

It has been mentioned long ago [30] and confirmed by the recent
studies [31-33] that the Local Group of galaxies is archetypical
for galactic systems on the spatial scale of $\sim$ 1 Mpc. The
group is dominated by the Milky Way and M31 located at about 0.7
Mpc from each other and moving toward each other with a relative
velocity $\sim$ 100 km/s. Other members of the group are the
Magellanic Clouds, the Triangulum galaxy and about four dozen
other dwarf galaxies. The group is $\simeq 1.5$ Mpc across and its
total mass is $M = 1.3 \pm 0.3\times 10^{12} M_{\odot}$ [25-28].

In the vicinity of the group, a two dozen galaxies move apart of
the group and form the local flow of expansion discovered by
Hubble in 1929. The flow is rather regular: it follows closely the
linear velocity-distance relation known as the Hubble law. The
rate of expansion (the local Hubble factor)is $H_{LG} = 72 \pm 6$
km/s/Mpc [25-28], which is almost exactly equal to the global
expansion rate measured by WMAP observations [8,9]  and close to
the expansion rate determined on the scale interval of 4-200 Mpc
[34-36]. The flow is `cool', and its velocity dispersion is less
than 20 km/s [25-28].

These data have been obtained in systematic observations of
distances and motions near the Local Group carried out over the
last eight years with the Hubble Space Telescope during more than
200 orbital periods [14-28]. High precision measurements were made
of the radial velocities (with 1-2 km/s accuracy) and distances
(8-10 \% accuracy) for about 200 galaxies of the Local Group and
neighbors from 0 to 7 Mpc from the group center. The 6-m BTA of
SAO and the Nordic Optical Telescope were also used, as well as
results from the KLUN-project [37-42].

The theory approach to the local flows of expansion developed by
us [10-13, 43-48] assumes that the flow structure and evolution is
controlled by the gravity of the central group and the antigravity
of the dark energy background in which all the galaxies of the
group and the flow are embedded.

Each galaxy of the flow is affected by the gravitational
attraction of the Local Group. Considering only the most important
dynamical factors, we may take the gravity field of the group as
nearly centrally-symmetric and static (this is a good
approximation to reality, as exact computer simulations prove
[44-46]. Then, according to Newtonian gravity, a galaxy is given
an acceleration
\be F_N = - GM/R^2,\ee
at its distance $R$ from the barycenter (which is the origin of
our reference frame).

The local antigravity is produced by the dark energy of vacuum
with the uniform local density $\bar \rho_V$. Then, according to
the `Einstein antigravity law', the dark energy produces
acceleration
\be F_E  = G 2\bar \rho_V (\frac{4 \pi}{3}R^3)/R^2 =  \frac{8
\pi}{3} G \bar \rho_V R, \ee \noindent where $-2 \bar \rho_V$ is
the local effective gravitating density of dark energy.  For
details see [12]. Eqs.2,3 describe the force field in the terms of
Newtonian mechanics; a General Relativity equivalent is given by
the static Schwarzschild-de Sitter space-time [12].

It is seen from Eqs.2 and 3 that the gravitational force ($\propto
1/R^2$) dominates over the antigravity force ($\propto R$) at
small distances where the acceleration is negative. At large
distances, antigravity dominates, and the acceleration is
positive.  Gravity and antigravity balance each other, so the
acceleration is zero, at the ``zero-gravity surface'' which has a
radius
\be R_V =  (\frac{3}{8\pi}M/\bar \rho_V)^{1/3}. \ee The
zero-gravity surface remains practically unchanged since the
formation of the Local group some 10-12 Gyr ago, as the computer
simulations [44-46] indicate.

The zero-gravity radius is a local spatial counterpart of the
``global'' redshift $z_V$: both indicate the gravity-antigravity
balance. However, there is a significant difference between the
global Friedmann theory and the local theory of Eqs.2-4. Indeed,
the global gravity field is uniform and time-dependent, while the
local field is non-uniform and static. Globally, the
gravity-antigravity balance takes place only at one proper-time
moment (at $z = z_V$) in the whole Universe. On the contrary, the
local gravity-antigravity balance exists since the formation of
the Local Group, but only at one distance ($R = R_V$).

The nearby expanding flow is made of dwarf galaxies.  It is
probable that the motions of these galaxies originate from the
early days of the Local Group when its major and minor galaxies
participated in violent non-linear dynamics with multiple
collisions and mergers. Our theory and computer simulations
[44-46] involve the concept of the ``Little Bang'' [49] as a model
for the origin of the local expansion component. The model shows
that some of the dwarf galaxies managed to escape from the
gravitational potential well of the Local Group after having
gained escape velocity from the non-stationary gravity field of
the forming group.

When the escaped galaxies occur beyond the zero-gravity surface
($R > R_V$), their motion is controlled mainly by the dark energy
antigravity and their trajectories are nearly radial there [12,
44-46]. The trend of the dynamical evolution controlled by dark
energy is seen from the fact that (as Eqs.2-4 show) at large
enough distances where antigravity dominates over gravity almost
completely, the velocities of the flow are accelerated and finally
they grow with time exponentially: $V \propto \exp [H_V t]$.
Because of this, the expansion flow acquires the linear
velocity-distance relation asymptotically:
\be V \rightarrow H_V R.  \ee \noindent
Here the value
\be H_V = (\frac{8 \pi G}{3} \bar \rho_V)^{1/2} \ee \noindent is
the universal expansion rate which is constant and determined by
the local dark energy density only [12].

If one assumes that the local density of dark energy is equal to
the global figure of Eq.1, the value of the universal rate may be
estimated as $H_V = 62 \pm 3$ km/s/Mpc. This is very close to the
observed expansion rates at distances 1-3 Mpc and 4-200 Mpc (see
above).

These considerations developed first in application to the close
vicinity of the Milky Way can be generalized and extended to other
local volumes on the spatial scale of a few Mpc. As we mentioned
above, observations indicate that other small groups and their
environment are similar to the local volume [30-33]. Computer
identified groups from observational galaxy catalogs [50] have
been shown to have an expanding population via a Doppler shift
number asymmetry relative to the brightest member. In addition,
large N-body $\Lambda$CDM cosmological simulations [51-56] show
that such a structure is rather typical for scales of a few Mpc
and more.

The structure with a massive galaxy group (or cluster) in its
center and a cool expansion outflow outside dominated by dark
energy seems to be a basic entity in the local structure of the
universe. We will refer to it as a {\it Hubble cell}.

\section{Cen A/M83 Hubble cell}

The Hubble cell of Cen A/M81 contains the galaxy group which is
called the Cen A/M83 complex [29] and the expansion outflow around
it. This is the second nearest Hubble cell to the Local Hubble
cell. A list of all known galaxies in a wide vicinity of a radius
about 4 Mpc around the barycenter of the group contains 87
objects; 38 of them have no velocity or/and distance estimates.
Most of the galaxies in the central volume of $\simeq 4$ Mpc
across form two families -- one, around the galaxy Cen A, includes
mostly ellipticals and the other, around the galaxy M83, includes
mostly spirals. The relative radial velocity of the centers of the
two families is near zero. Their 3D separation is 1.3-2 Mpc.

The mass of the Cen A family (the central galaxy including) is
estimated as $M_{CenA} \simeq (6-8)\times 10^{12} M_{\odot}$. The
mass of the other family is about an order of magnitude less:
$M_{M83} \simeq 1 \times 10^{12} M_{\odot}$. Thus, the total mass
of the complex is practically the mass of the giant elliptical
galaxy Cen A [29] and its surrounding.

Observational data [29] on the velocities and distances in the Cen
A/M83 Hubble cell are presented in the Hubble diagram of Fig.1.
The velocities and distances are given relative to the Cen A
galaxy. The distances are determined with considerable errors
which are shown in Fig.1 (without the error of the position of the
central Cen A galaxy.) The accuracy of the distance determination
is significantly lower there than in the Local Hubble cell where
the error is typically about 10\%.

The flow of expansion is clearly seen in Fig.1 at the distances
2-3 Mpc from the group barycenter (coincident with the position of
the giant galaxy Cen A). The total number of the receding galaxies
at the distances $< 4$ Mpc is 21. The flow reveals the linear
velocity-distance relation (the Hubble law) with the expansion
rate about 72 km/s/Mpc. The flow is cool enough: the velocity
dispersion is about 30 km/s. As this value is affected
significantly by the distance determination errors, the true value
is still lower.

Using the general relation of Eq.4, one may estimate the
zero-gravity radius for the Cen A/M83 group. With the mass of the
group assumed as $M \simeq M_{CenA} = 7\times 10^{12} M_{\odot}$
and the dark energy density $\rho_V = 0.75 \times 10^{-29} g/
cm^{3}$, one finds: $R_V \simeq 2$ Mpc. According to the
considerations of the section above, the members of the galaxy
group (complex) must be located within the zero-gravity surface,
and this is really so, as one can see from Fig.1. In particular,
the galaxy M83, the second largest member of the group, is (most
probably) located at the distance $R < R_V$. All the galaxies with
negative velocities are also within the zero-gravity sphere. The
considerations of Sec.2 indicate as well that the flow of
expansion is expected to approach the linear velocity-distance
relation outside the zero-gravity surface where the antigravity of
the dark energy background dominates over the gravity of the group
matter. Indeed, Fig.1 shows that this relation emerges starting
from the distance $R \simeq R_V \simeq 2$ Mpc. All the galaxies at
the distances $R > R_V$ move apart of the group; there is no
infall (negative velocities) on the group at $R > R_V$.

The theory of Sec.2 makes a definite prediction: the expansion
rate at distances $R > R_V$ must be near the universal expansion
rate $H_V \simeq 62$ km/s/Mpc. The data above agree with this
prediction within 15 \% accuracy.

The theory makes also another specific prediction. It follows from
Eqs.2-6 that at distances $R > R_V$, the velocities of the local
expansion flow must be not less than a minimal velocity $V_{esc}$
[12]:
\be V_{esc} = (\frac{2G M}{R_V})^{1/2} (\frac{R}{R_V})^{1/2} [1 +
\frac{1}{2} (\frac{R}{R_V})^3 - \frac{3}{2}
(\frac{R}{R_V})]^{1/2}. \ee

The minimal velocity corresponds to the minimal total mechanical
energy,
\be E_{esc} = - \frac{3}{2} \frac{GM}{R_V}, \ee \noindent needed
for a particle to escape from the gravitational potential well of
the group. Due to the antigravity of dark energy, this energy is
negative. The velocity $V_{esc} = 0$ at $R = R_V$.

Actually, this prediction may serve as a critical test for the
theory. In Fig.1, the minimal velocity $V_{esc}$ is showed by a
bold curve. The curve is determined by two parameters which are
the group mass $M$ and the dark energy density $\rho_V$; it is
assumed in Fig.1 that $M = 7 \times 10^{12} M_{\odot}$ and the
local dark energy density is equal the global dark energy density
(see Eq.1). It is seen in the figure that all the galaxies at the
distances $R > R_V$ obey this prediction: none of 21 galaxies of
the flow at $ R > R_V$ lies below the critical line. In this way,
the data of Fig.1 indicate that the theory passes the test. It is
even more important that the data are in agreement with the two
basic parameters of the theory which are the group mass and the
dark energy density. This means that the local density of dark
energy in the Cen A/M83 Hubble cell is near or exactly equal to
the global figure of Eq.1.

As in the case of the Local Hubble cell [12,13], one may follow
the logic of the global determination of the dark energy density
(see Sec.1) to estimate independently the value of the local dark
energy density in the Cen A/M83 Hubble cell. For this goal, one
needs first to determine the zero-gravity radius $R_V$. Basing on
the theory considerations of Sec.2, one may robustly restrict the
value of $R_V$ with the use of the diagram of Fig.1. Indeed, since
the zero-gravity surface lies outside the group (complex) volume,
it should be that $R_V > 2 $ Mpc. On the other hand, the fact that
the linear velocity-distance relation is seen from a distance of
about, say, 3 Mpc suggests that $R_V < 3$ Mpc. If so, Eq.4 leads
directly to the robust upper (from $R > 2$ Mpc) and lower ($R < 3$
Mpc) limits to the local dark energy density:
\be (0.1 \pm 0.1) < \rho_V < (1 \pm 0.1) \times 10^{-29} \;\;
g/cm^3. \ee \noindent (Here the measured value of the mass of the
complex is also used.)

The lower limit in Eq.9 is most significant. It means that the
dark energy does exist in the Hubble cell. In combination, both
limits imply that the value of the local dark energy density is
near the value of the global dark energy density, or may be
exactly equal to it.

In addition, the structure of the flow follows the trend of the
minimal velocity: the linear regression line of the flow (the thin
line in Fig.1) is nearly parallel to the minimal velocity curve,
at $R > R_V$. It may easily be seen from the theory of Sec.2 that
in the limit of large distances, the minimal velocity and the real
velocity of the flow galaxies have a common asymptotics, $V
\propto V_{esc} \propto H_V R$, independently on the initial
conditions of the galaxy motion.

For a comparison, a similar minimal escape velocity, $(\frac{2G
M}{R_V})^{1/2} (\frac{R}{R_V})^{1/2}$ is showed in Fig.1 for a
``no-vacuum model'' with zero dark energy density (dashed line).
The real flow is obviously ignores the trend of the minimal
velocity in this case: the velocities of the flow grow with
distance, while the minimal velocity decreases. It is seen also
that 19 galaxies  of the flow at $R > R_V$ violate obviously the
no-vacuum model: they are located below the dashed line. This
comparison is clearly in favor of the vacuum model and against the
model with no dark energy.

\section{Conclusions}

The Hubble cell is a notion in cosmology which appears with the
new understanding of the local flows of expansion suggested in our
works [10-13, 44-48]. The archetypical example of the Hubble cell
is the Local cell which includes the Local Group and the cool
expansion outflow around it. The major new physics discovered in
the Hubble cell is the presence of dark energy and its domination
in the dynamics of the expansion flows. The basic physical
quantity which is characteristic for the cell is the zero-gravity
radius $R_V$ introduced in [10-13]. For the Local Hubble cell,
$R_V = 1.2 \pm 0.1$ Mpc. The central group of the Hubble cell is
located within the zero-gravity surface ($R < R_V$) and controlled
mostly by the gravity of the group matter. The cool expansion
outflow develops outside the surface ($R > R_V$), and its
structure and evolution are determined by the antigravity of the
dark energy background. The surface is nearly spherical and it is
nearly unchanged for the life-time of the Hubble cell.

A simple theory of the Hubble cell is based on the description of
the gravity/antigravity interplay in terms of the Newtonian
mecanics. The theory incorporates the concept of the Little Bang
[49] according to which the expansion flow is formed by the dwarf
galaxies that escape from the central group. The antigravity of
the dark energy background makes the escape energy barrier lower,
and in this way, it stimulates the process. The theory
demonstrates that the major trend of the flow dynamical evolution
is the development from chaos to order under the action of the
antigravity of the perfectly uniform dark energy background. The
initially chaotic ensemble of the trajectories of the escaped
members of the group gains a regular structure in the phase space
of the cell: 1) the trajectories tend to the linear
velocity-distance relation; 2) they also tend to the universal
expansion rate $H_V \simeq 62$ km/s/Mpc which is determined by the
dark energy density alone and 3) the velocity dispersion in the
flow decreases with time due to ``vacuum cooling'' which is much
more effective than the adiabatic cooling. This evolutionary trend
means that the Hubble cell are stable structures -- any changes in
their dynamics and kinematics lead only to a higher degree of
order and regularity in them.

The Cen A/M83 Hubble cell ($R_V \simeq 2$ Mpc) studied in the
present paper is a close analogue of the Local Hubble cell. The
mass of this cell is considerably larger, and the mass is mainly
associated with the central giant galaxy Cen A. Because of this,
the approximation of the static spherical symmetry works in this
case even better than in the case of the Local Hubble cell. The
third example of the Hubble cell is the M81 galaxy group with its
expansion outflow [57]. The gravity of this cell is dominated by
the giant galaxy M81 (its mass is near the mass of the Local Group
and $R_V \simeq 1.2$ Mpc), and so the theory approximation above
is practically exact in this case. On a larger scale, the Virgo
cluster and the Coma cluster with the expansion outflows around
each of them ($R_V \simeq 8$ and $15$ Mpc, respectively) may be
treated as Hubble cells which extend to the distances $simeq 15 $
and $30$ Mpc.

Hubble cells acquire a role of the building blocks of the local
universe. A world wide network of the Hubble cells may cover the
observed space almost entirely. As a result, the structure of the
universe within the cosmic cell of uniformity (scales less than
100-300 Mpc) proves to be more regular and better organized than
it may be seen at the first glance. Indeed, the Hubble cells of
various scales are all well ordered kinematically. Their
zero-gravity surfaces are almost spherical with their radii almost
constant in time during the last 10-12 Gyr. The overall dark
energy domination makes the universe almost uniform on the scales
exceeding the sizes of the zero-gravity radii which are 1-10 Mpc.
On the same scales, the expansion rate is near the universal value
$H_V$ due to the dark energy dominance in the cell outflows. As is
well known, the regularity of the quiet expansion flow -- in spite
of matter high irregularities -- has been considered as a big
mystery for decades [36, 58-60].

To conclude, modern cosmology emerged from the discovery of dark
energy [1,2] is rich of new exciting implications that are due to
the realization that dark energy rules not only the universe as a
whole, but also our close galaxy environment [10-13]. The concept
of the new type of astronomical objects -- the Hubble cells -- is
one of these implications. These objects have a stable regular
structure, they are omnipresent and introduce simplicity and order
to the grand design of the universe on a wide range of spatial
scales.

The work of A.C., V.D. and L.D. was partly supported by a RFBR
grant 06-02-16366.

\section*{References}

\noindent [1] A.G. Riess, A.V. Filippenko, P. Challis et al. AJ,
{\bf }, 1009 (1998).

\noindent [2] S. Perlmuter, G. Aldering, G. Goldhaber G. et al.
ApJ, {\bf 517}, 565 (1999).

\noindent [3] A.G. Riess et al., ApJ, {\bf 627}, 579 (2005).

\noindent [4] A.G. Riess et al., ApJ, {\bf 607}, 665 (2004).

\noindent [5] M. Sullivan et al. AJ, {\bf 131}, 960 (2006).

\noindent [6] P. Astier et al. A\&A, {\bf 447}, 31 (2006).

\noindent [7] E.B. Gliner.  Sov.Phys. JETP, {\bf 22}, 378 (1966).

\noindent [8] D.N. Spergel et al. ApJS, {\bf 148}, 175 (2003).

\noindent [9] D.N. Spergel et al. astro-ph/0603449 (2006).

\noindent [10]  A.D. Chernin, P. Teerikorpi, Yu.V. Baryshev. Adv.
Space Res., {\bf 31}, 459, (2003).

\noindent [11] A.D. Chernin. Physics-Uspekhi, {\bf 44}, 1099
(2001).

\noindent [12] A.D. Chernin, P. Teerikorpi, Yu.V. Baryshev. A\& A,
{\bf 456}, 13 (2006).

\noindent [13] A.D. Chernin, I.D. Karachentsev, P. Teerikorpi et
al. A\&A (2007) (in press).

\noindent [14] I.D. Karachentsev et al. A\&A, {\bf 389}, 812
(2002).

\noindent [15] I.D. Karachentsev, O.G. Kashibadze. Astrofizika
{\bf 49}, 5 (2006).

\noindent [16] I.D. Karachentsev et al. astro-ph/0603091 (2006).

\noindent [17] I.D. Karachentsev, M.E. Sharina, E.K. Grebel et al.
A\&A, {\bf 352}, 399 (1999).

\noindent [18] A.E. Dolphin, L.M. Makarova, I.D. Karachentsev et
al. MNRAS, {\bf 324}, 249 (2001).

\noindent [19] I.D. Karachentsev, M.E. Sharina, A.E. Dolphin et
al. A\&A, {\bf 379}, 407 (2001).

\noindent [20] I.D. Karachentsev, A.E. Dolphin, D. Geisler et al.
A\&A, {\bf 383}, 125 (2002).

\noindent [21]  I.D. Karachentsev, M.E. Sharina, A.E. Dolphin et
al. A\&A, {\bf 385}, 21 (2001).

\noindent [22] I.D. Karachentsev, M.E. Sharina, D.I. Makarov et
al. A\&A, {\bf 389}, 812 (2002).

\noindent [23] I.D. Karachentsev, D.I. Makarov, M.E. Sharina et
al. A\&A, {\bf 398}, 479 (2003).

\noindent [24] I.D. Karachentsev, E.K. Grebel, M.E. Sharina et al.
A\&A, {\bf 404}, 93 (2003).

\noindent [25]  I.D. Karachentsev, M.E. Sharina, A.E. Dolphin,
E.K. Grebel. A\&A, {\bf 408}, 111 (2003).

\noindent [26] I.D. Karachentsev, V.E. Karachentseva, W.K.
Huchtmeier, D.I. Makarov. AJ, {\bf 127}, 2031 (2004).

\noindent [27] I.D. Karachentsev. AJ, {\bf 129}, 178 (2005).

\noindent [28] I.D. Karachentsev, A.E. Dolphin, R.B. Tully. AJ,
{\bf 131}, 1361 (2006).

\noindent [29] I.D. Karachentsev, R.B. Tully, A.E. Dolphin et al.
AJ, {\bf 133}, 504 (2007).

\noindent [30] E. Hubble. {\it The Realm of the Nebulae},  Oxford
Univ. Press, Oxford (1936).

\noindent [31] S. van den Bergh. astro-ph/0305042 (2003).

\noindent [32] S. van den Bergh. AJ, {\bf 124}, 782 (2002).

\noindent [33] S van den Bergh. ApJ, {\bf 559}, L113 (2001).

\noindent [34] F. Thim, G. Tammann, A. Saha et al. ApJ, {\bf 590},
256 (2003).

\noindent [35] A. Sandage, G.A. Tamman, B. Reindl. A\&A, {\bf
424}, 43 (2004).

\noindent [36] A. Sandage, G.A. Tamman, A. Saha et al. ApJ, {\bf
653}, 843 (2006).

\noindent [37] R. Rekola, M.G. Richer, M.L. McCall et al. MNRAS,
361, 330 (2005).

\noindent [38] T. Ekholm, P. Teerikorpi, G. Theureau et al. A\&A,
{\bf 347}, 99 (1999).

\noindent [39] P. Teerikorpi, M. Hanski, G. Theureau. et al. A\&A,
{\bf 334}, 395 (1998).

\noindent [40] G. Theureau, M. Hanski, T. Ekholm et al.  A\&A,
{\bf 322}, 730 (1997).

\noindent [41] T. Ekholm, Yu. Baryshev, P. Teerikorpi et al. A\&A,
{\bf 368}, 17 (2001).

\noindent [42] G. Paturel, P. Teerikorpi. A\&A, {\bf 443}, 883
(2005).

\noindent [43] Yu. Baryshev, A. Chernin, P. Teerikorpi. A\&A, {\bf
378}, 729 (2001).

\noindent [44] A.D. Chernin, I.D. Karachentsev, M.J. Valtonen et
al. astro-ph//057364 (2005).

\noindent [45] A.D. Chernin, I.D. Karachentsev, M.J. Valtonen et
al. A\&A, {\bf 415}, 19 (2004).

\noindent [46] V.P. Dolgachev, L.M. Domozhilova, A.D. Chernin.
Astr. Rep., {\bf 47}, 728 (2003).

\noindent [47] I.D. Karachentsev, A.D. Chernin, P. Teerikorpi.
Astrofizika {\bf 46}, 491 (2003).

\noindent [48] P. Teerikorpi, A.D. Chernin, Yu.V. Baryshev. A\&A,
{\bf 440}, 791 (2005).

\noindent [49] G.G. Byrd, M.J. Valtonen, M. McCall, K. Innanen.
AJ, {\bf 107}, 2055 (1994).

\noindent [50]  M.J. Valtonen, G.G. Byrd. ApJ, {\bf 303}, 523
(1986).

\noindent [51] K. Nagamine, R. Cen, J.P. Ostriker. Bul. Amer.
Astron. Soc., {\bf 31}, 1393 (1999).

\noindent [52] K. Nagamine, J.P. Ostriker, R. Cen. ApJ. {\bf 553},
513 (1991).

\noindent [53] J.P. Ostriker, Y. Suto. ApJ, {\bf 348}, 378 (1990).

\noindent [54] Y. Suto, R. Cen, J.P. Ostriker. ApJ, {\bf 395}, 1
(1992).

\noindent [55]  V.A. Strauss, R. Cen, J.P. Ostriker. ApJ, {\bf
408}, 389 (1993).

\noindent [56] A.V. Macci\`{o}, F. Governato, G. Horellou. MNRAS,
{\bf 359}, 941 (2005).

\noindent [57] A.D. Chernin, I.D. Karachentsev, P. Teerikorpi et
al. (2007) (in preparation).

\noindent [58] A. Sandage. ApJ, {\bf 307}, 1 (1986).

\noindent [59] A. Sandage. ApJ, {\bf 527}, 479 (1999).

\noindent [60] A. Sandage et al. ApJ, {\bf 172}, 253 (1972).

\begin{figure}
\centerline{\includegraphics{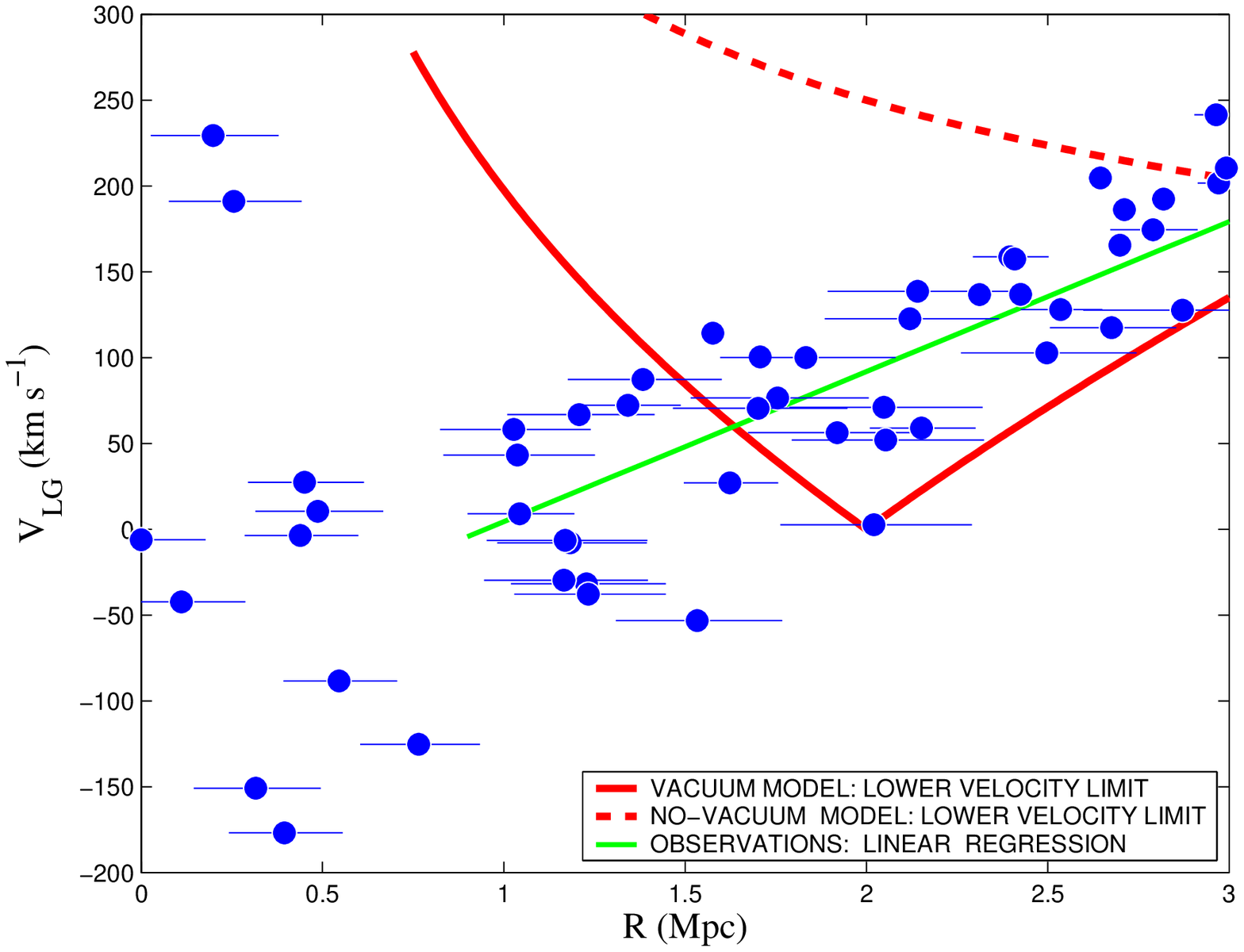}}
\caption{The Hubble velocity-distance diagram for the Cen A/M83
Hubble cell based on the data [29]. The velocities and distances
are given relative to the Cen A galaxy. The galaxies of the Cen
A/M83 complex are located within the area of 4 Mpc across. The
flow of expansion starts in the outskirts of the group; all the
galaxies at distances $R > 2$ Mpc move apart of the group
(positive velocities). The flow reveals the linear
velocity-distance relation  known as the Hubble law at $R \ge 2$
Mpc (see also the text).}
\end{figure}

\end{document}